\documentclass[hyper]{JINST}
\usepackage[utf8]{inputenc}
\usepackage{indentfirst}

\title{Demonstrating a directional detector based on neon for characterizing high energy neutrons}
\author{A. Hexley$^{1}$, M.H. Moulai$^{1}$, J. Spitz$^{2}$, J.M. Conrad$^1$\\
$^{1}$Massachusetts Institute of Technology, Cambridge, MA, 02139, USA \\
$^{2}$University of Michigan, Ann Arbor, MI, 48109, USA}

\date{July 2015}

\usepackage{graphicx}

\abstract{MITPC is a gas-based time projection chamber used for detecting fast, MeV-scale neutrons. The standard version of the detector relies on a mixture of 600~torr gas composed of 87.5\% $^4$He and 12.5\% CF$_4$ for precisely measuring the energy and direction of neutron-induced nuclear recoils. We describe studies performed with a prototype detector investigating the use of Ne, as a replacement for $^4$He, in the gas mixture. Our discussion focuses on the advantages of Ne as the fast neutron target for high energy neutron events ($\lesssim$100 MeV) and a demonstration that the mixture will be effective for this event class. We find that the achievable gain and transverse diffusion of drifting electrons in the Ne mixture are acceptable and that the detector uptime lost due to voltage breakdowns in the amplification plane is negligible, compared to $\sim 20\%$ with the $^4$He mixture.}

\begin{document}

\maketitle

\section{Introduction}

The MIT and Michigan Time Projection Chamber (MITPC) is a directional fast neutron detector which relies on a CCD camera and the TPC technique for creating a three dimensional image of neutron-induced nuclear recoils within a gaseous volume. The detector ran for over a year at Double Chooz, a reactor-based neutrino experiment in France, under the name ``Double Chooz Time Projection Chamber'' (DCTPC), and is now installed in the Booster Neutrino Beamline (BNB) in the SciBooNE Enclosure at Fermi National Accelerator Laboratory (FNAL), where it will begin taking beam data in Fall 2015. Fast, MeV-scale neutrons represent a significant background for, among others, neutrinoless double beta decay, solar neutrino, reactor neutrino, coherent neutrino, and WIMP measurements. MITPC is designed to characterize neutron backgrounds, measuring rate as a function of energy, direction, depth underground, and rainfall~\cite{prototype}. MITPC can also be used to study fast neutron backgrounds in beam-based neutrino experiments. These neutrons, produced via neutrino interactions with materials in or around the detectors, are a background for a number of important neutrino processes, including low energy charged current and neutral current elastic scattering~\cite{ncelastic}. Further, understanding neutrino-induced neutrons as a function of neutrino energy, $Q^2$, etc. is important for future oscillation and neutrino cross section measurements (see, e.g., Ref.~\cite{huber}). Neutrons usually represent missing energy in neutrino event reconstruction and it is and will be important to quantify this effect for the precise neutrino calorimetry that current and future cross section and oscillation measurements depend on. A number of experiments will study this topic specifically~\cite{annie,sbnd,microboone}.  Informing these experiments, especially those using the BNB, about the energy and directionality of neutrino- and beam-related (``skyshine'') neutrons is MITPC's main goal.  

\section{Configuration of the Detector}

\begin{figure}[t]
\begin{center}
\includegraphics[scale=0.41]{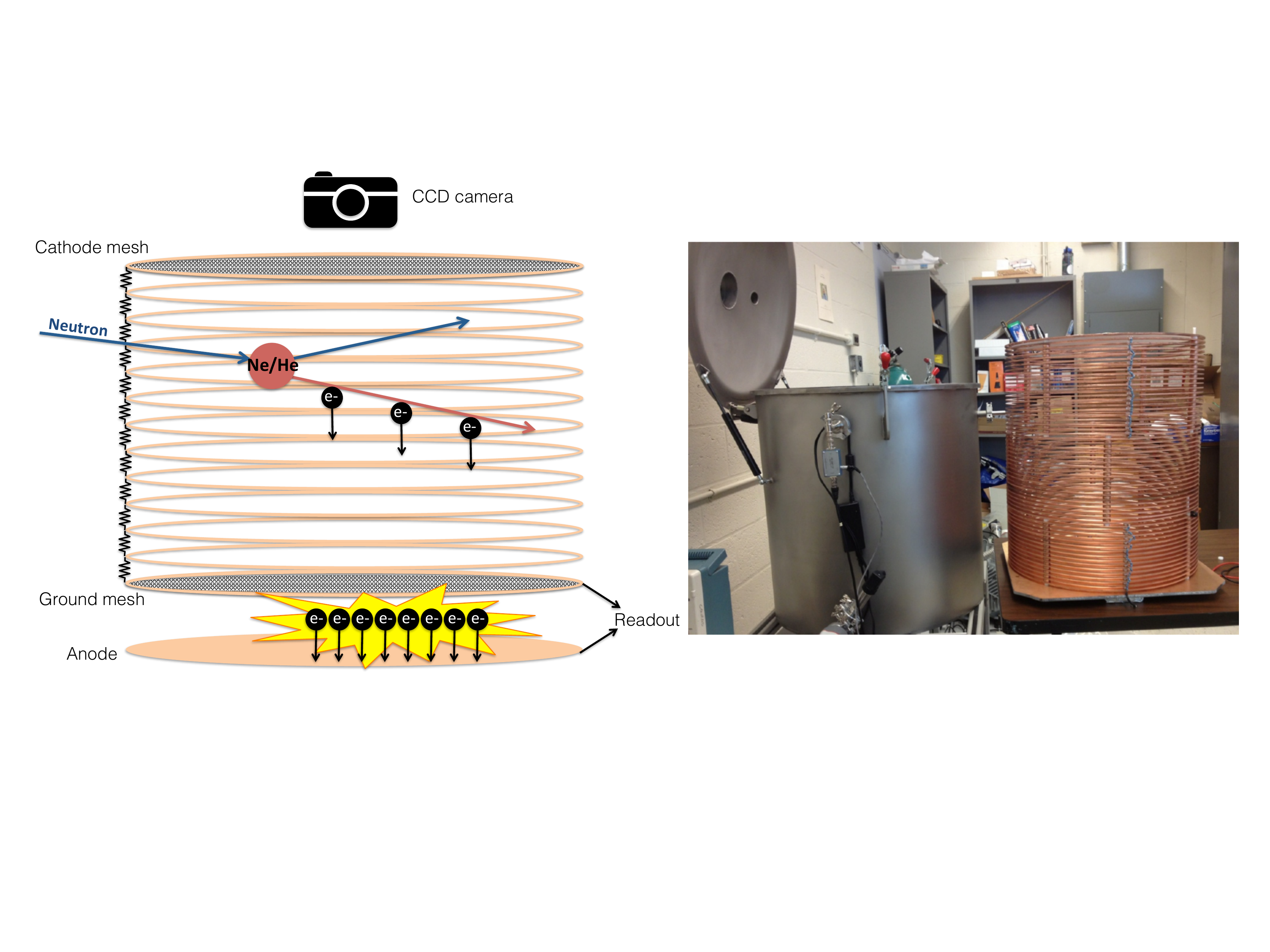} 
\vspace{-3.2cm}
\caption{Left: a schematic of a neutron event occurring inside the detector. Right: the full size detector's field cage next to the vacuum vessel. \label{schematic}}
\end{center}
\end{figure}

Two DCTPC detectors were constructed for use at Double Chooz: a prototype detector (2.8~L)~\cite{prototype} and a full-size detector (originally 60~L, now 40~L). A paper describing the results of the DCTPC run at Double Chooz is in preparation. MITPC is the reconfigured 40~L detector installed at FNAL. Here, we provide a brief description of MITPC to facilitate discussions of the Ne measurements performed with the prototype detector.

The TPC consists of a copper field cage, a cathode mesh, and an amplification plane placed in a vessel filled with gas. The field cage is located directly above the amplification plane, which consists of a copper anode plate and stainless-steel ground mesh separated by 0.4~mm (shown in Figure~\ref{schematic}). The gas composition is 87.5$\%$ of either $^4$He (DCTPC at Double Chooz) or Ne (MITPC at FNAL), and 12.5$\%$ CF$_4$. The CF$_4$ acts as the scintillator and quencher for the electron avalanche, while the $^4$He or Ne acts as the primary neutron target. The gas is held at 600 torr.

The amplification plane is imaged by a CCD camera (U6 model made by Apogee Instruments Inc., featuring a Kodak KAF-1001E 1024x1024 pixel CCD chip) and charge is read out from both the anode and ground mesh. In the prototype detector used for the Ne studies, the field cage height (maximum drift distance) is 10~cm, the amplification plane is 24.7~cm in diameter, and the CCD camera images an area of 16.7~cm x 16.7~cm in the center of the amplification plane. The reconstruction of events is described below and diagrammed in Figure~\ref{schematic}.

When a fast neutron enters the detector and interacts with a target nucleus, the nucleus recoils and ionizes the surrounding gas, leaving a track in its wake. Electrons drift through the 150~V/cm electric field of the field cage between the cathode mesh and amplification plane, until they reach the latter. The electrons enter the amplification plane, a high field ($\sim1750~\mathrm{V/mm}$) region between the ground mesh and the anode, and rapidly accelerate to create an ionization avalanche in the region of the track. The cascade produces a light signal that is imaged by the CCD and the charge in time is simultaneously collected by the anode and mesh planes via a set of preamplifiers.

\begin{figure}[t]
\begin{center}
\includegraphics[scale=0.35]{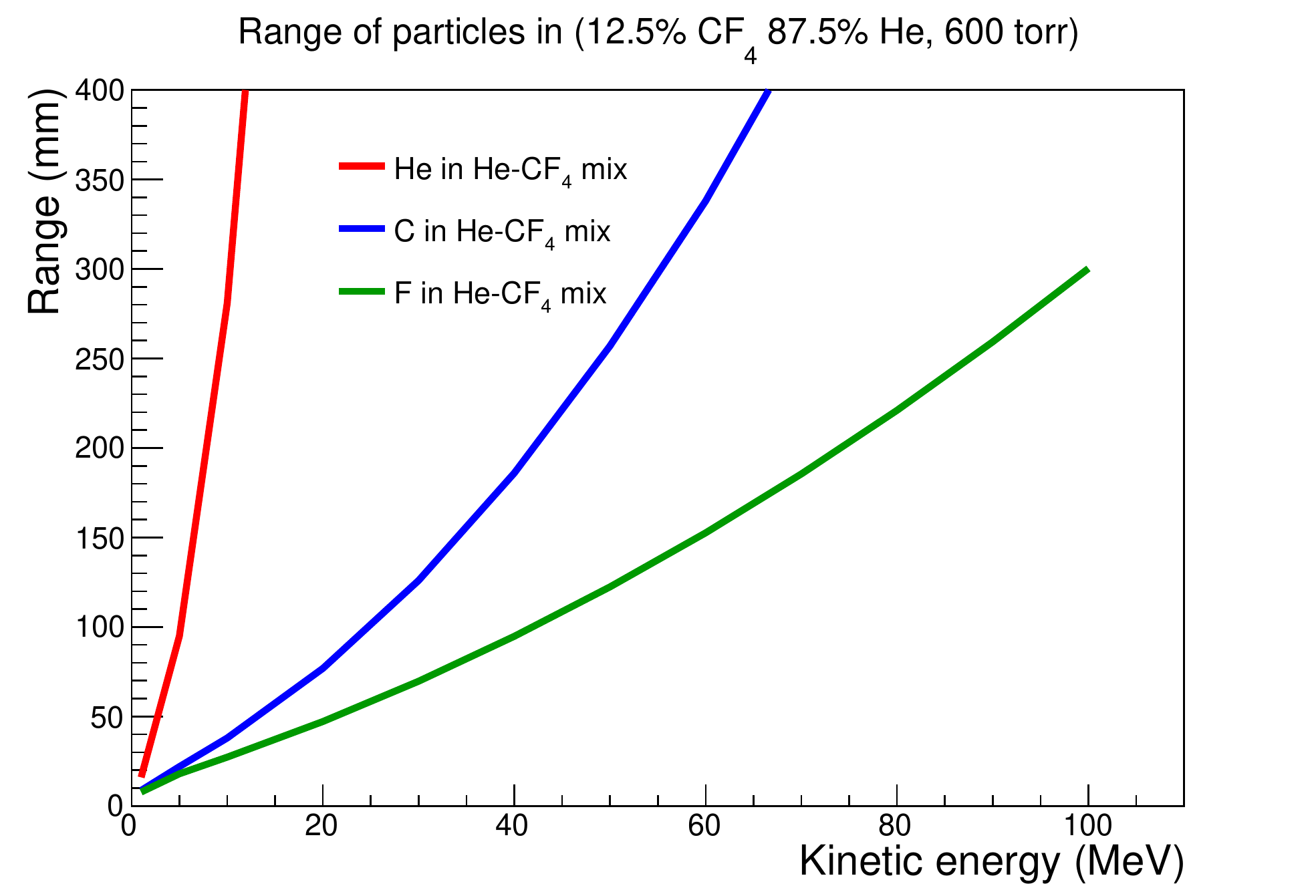}
\includegraphics[scale=0.35]{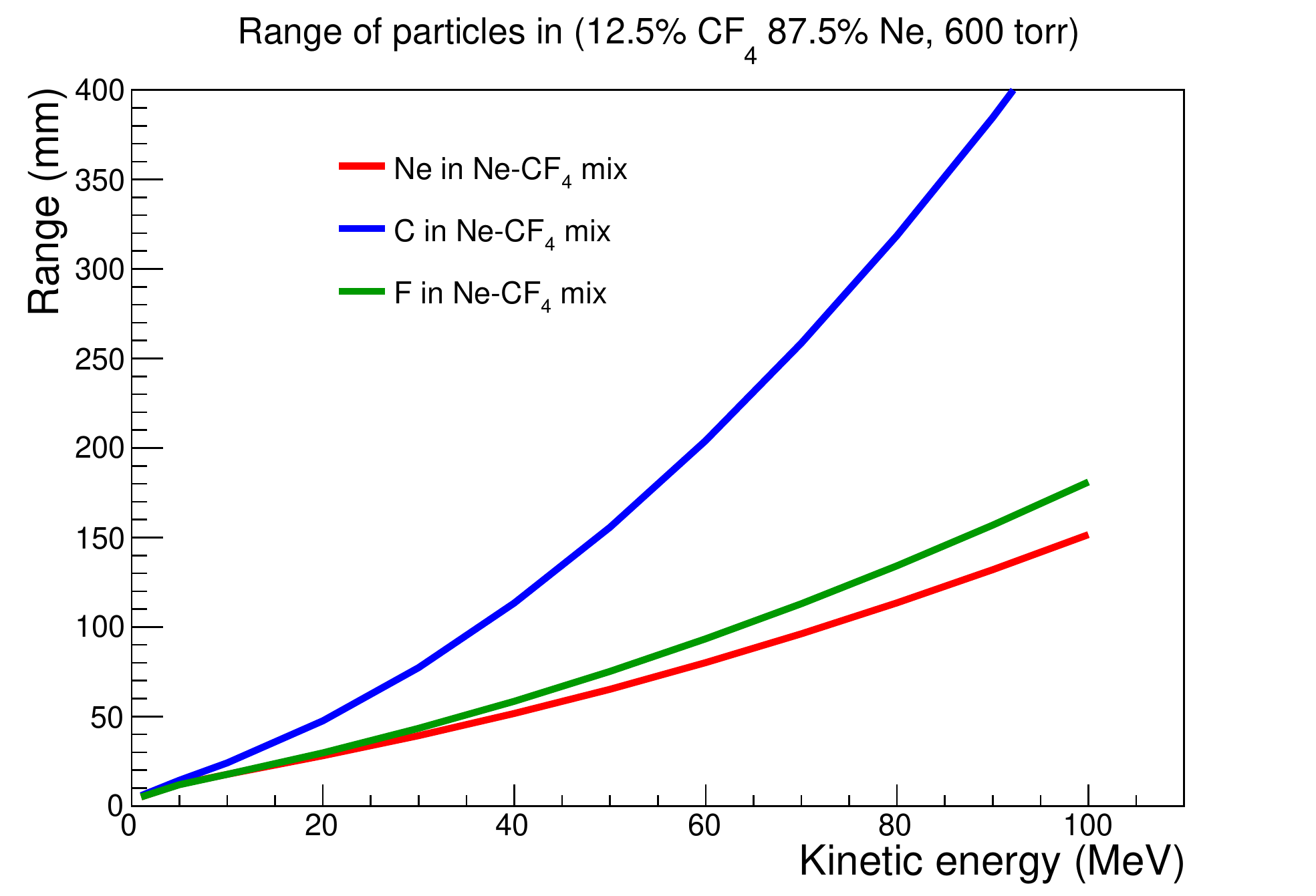}
\caption{The relationship between range and kinetic energy for nuclei travelling through the He-based mixture (left) and Ne-based mixture (right), according to the SRIM software~\cite{srim}.\label{srim}}%Cross sections are scaled by the concentration of gas.\label{srim}}
\end{center}
\end{figure}

A coincident signal on the three readouts (CCD, anode, and mesh) is considered an event. The 3D length of the event track is reconstructed by timing the difference between the start and end of the track arriving at the ground mesh. The vertical length of the track is found after combining this information with the known drift velocity in the chamber. The length of the track on the CCD readout, the ``CCD range'', is the 2D horizontal length of the track. Using the SRIM software package, the results of which are shown in Figure~\ref{srim}, we reconstruct the energy of the track based on its 3D length~\cite{srim}. For reference, the energy resolution in the $^4$He mixture configuration is about 2$\%$ at a nuclear recoil energy of 5~MeV. As we are only able to fully reconstruct events that are fully contained inside the detector, any track that crosses the edge of the CCD imaging area is rejected from consideration.

\begin{figure}[t]
\begin{center}
\includegraphics[scale=0.35]{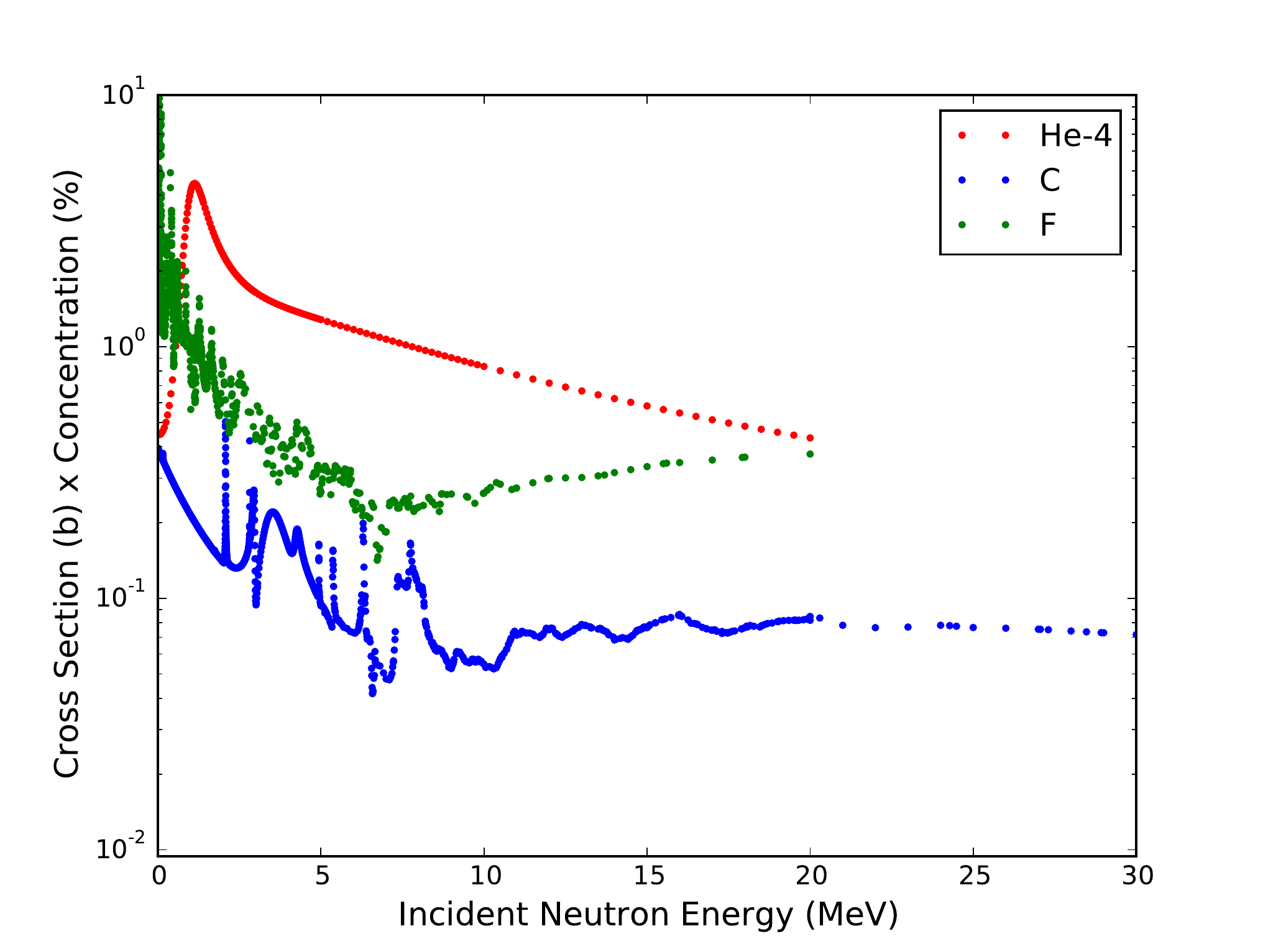}
\includegraphics[scale=0.35]{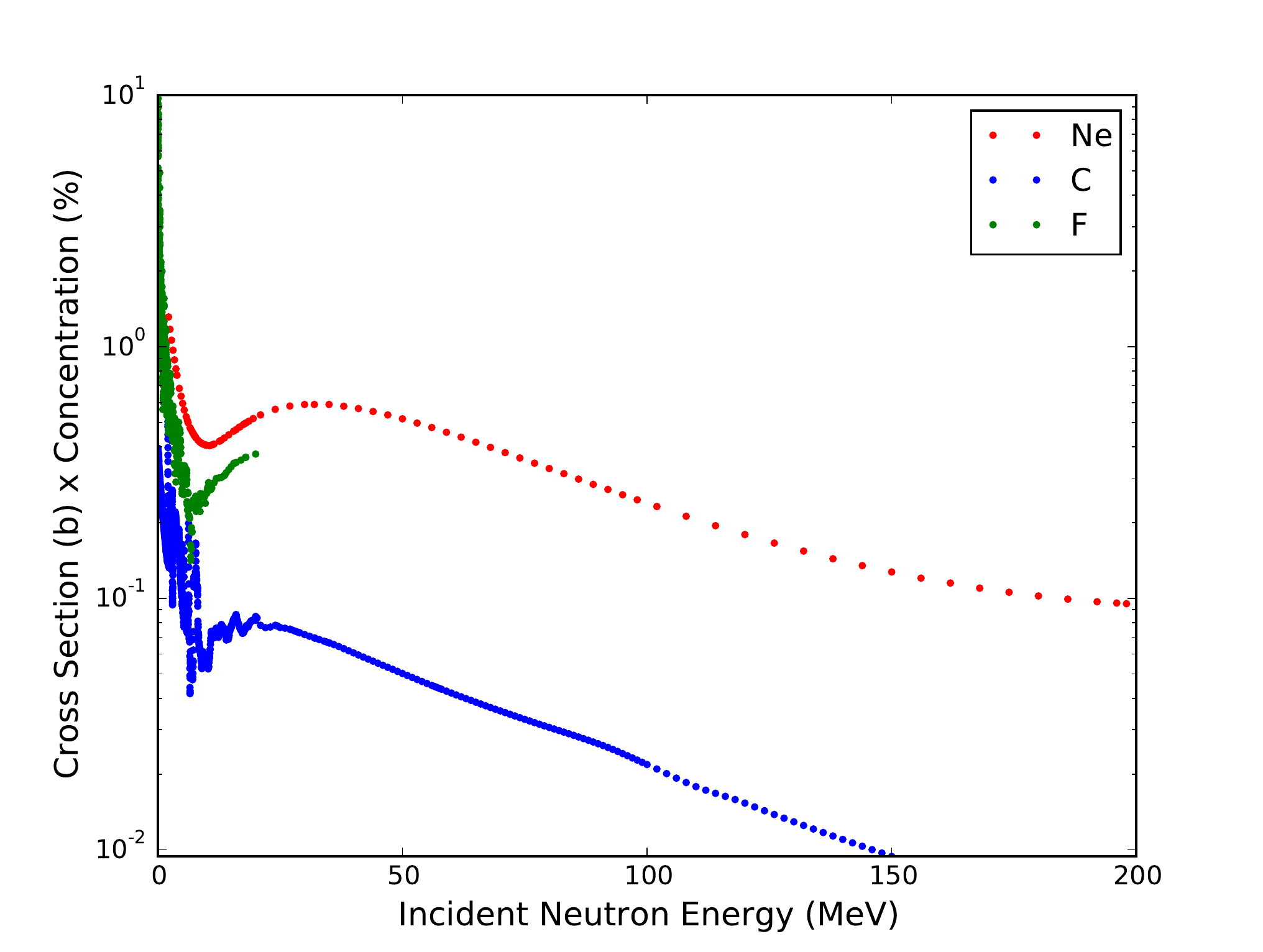}
\caption{The neutron elastic interaction cross section (weighted by concentration in the gas mixture employed) versus neutron energy for the relevant target nuclei and energy ranges in the $^4$He gas mixture (left) and Ne gas mixture (right). \label{xsec}}
\end{center}
\end{figure}

The energy loss profile of a track can help in identifying the recoiling nucleus, but it is important to understand the expected neutron event rate on each target in the gas mixtures considered. The neutron elastic interaction cross sections (weighted by concentration in the gas mixture) as a function of incident neutron energy for $^4$He, Ne, C, and F nuclei are shown in Figure~\ref{xsec}~\cite{xsection,xsection2}. The neutron-Ne elastic cross section surpasses $^4$He at about 15~MeV. In general, $^4$He/Ne dominates the interaction rate, as compared to CF$_4$, across the energy region of interest. The expected event rate on a $^4$He target is a factor of at least 2.5 greater than CF$_4$ in the neutron energy range 1--10 MeV. The $^4$He/CF$_4$ ratio begins to decrease after 10 MeV but is still at 1.4 at 15 MeV. Similarly, the Ne rate dominates over C; the ratio of Ne/C is $\sim$10 across the neutron energy range 50--150 MeV. There is no F data available in this energy region.

\section{Neon Studies}

\subsection{Motivation}
    The neutrons measured by DCTPC at Double Chooz originated from cosmic ray and radioactive sources. DCTPC was therefore tuned to accept neutron-induced nuclear recoils in the energy range 0.2--20 MeV. The neutrons of most interest at FNAL, however, are beam-induced neutrons which tend to be more energetic. The expected neutrino-induced neutron energy in the BNB ($<E_\nu>= 0.8~\mathrm{GeV}$) was studied using a Geant4- and GENIE-based simulation~\cite{geant4,genie} and found to be $\lesssim$400 MeV. In this higher energy range, most relevant for the BNB, the large majority of neutron-induced $^4$He nucleus recoils are not expected to be fully contained. The SRIM software package~\cite{srim} results for the expected range as a function of nuclear recoil energy on the relevant nuclei are shown in Figure~\ref{srim}. To fully contain a larger fraction of event tracks, we must use an element heavier than $^4$He, which will tend to travel a shorter distance in the gas medium. We therefore have turned to heavier noble gases to act as the fast neutron target for MITPC running at FNAL.

Signal-to-noise, especially in consideration of the CCD signal, is an important aspect of this detection technology. Electron diffusion, drift velocity, amplification plane gain, impurity content, recombination, and electron attachment can all affect signal-to-noise. While each of these is relevant for the Ne study outlined here, our concern, and therefore our focus, has been on electron diffusion and the resulting potential signal-to-noise decrease. The relevant literature suggests that electrons diffuse significantly more in noble gases heavier than $^4$He~\cite{noble_gas}, since the momentum transfer cross section is lower for these heavy atoms. Importantly, however, mixing noble elements with CF$_{4}$ is known to improve diffusion significantly (see, e.g., Ref.~\cite{hide}). Increased transverse diffusion reduces a CCD track's ``brightness'' (signal-to-noise ratio, nominally at $\sim$10:1, with the CCD readout noise of $\sim$10 e$^-$ RMS at 1~MHz dominating the sources of image noise) and can significantly jeopardize detector efficiency and energy resolution. Among the noble elements, Ne is expected to have the least diffusion after $^4$He, so we chose it for our initial studies in searching for an adequate heavy target for the characterization of high energy neutrons. 

We ran tests comparing a Ne/CF$_4$ mixture to a $^4$He/CF$_4$ mixture using the prototype detector at MIT. The objectives of our study were threefold: to study the diffusion in Ne, to determine whether we could obtain sufficient gain for high efficiency event reconstruction, and to test the spark rate in the amplification plane. An $^{241}$Am alpha source (4.4~MeV) was inserted into the center of the prototype detector 6.2~cm from the amplification plane. We ran tests with both Ne and $^4$He using a 600~torr gas mixture with a concentration fraction of 87.5$\%$ fast neutron target ($^4$He or Ne) and 12.5$\%$ CF$_4$.

\begin{figure}[t]
\begin{center}
\includegraphics[scale=0.4]{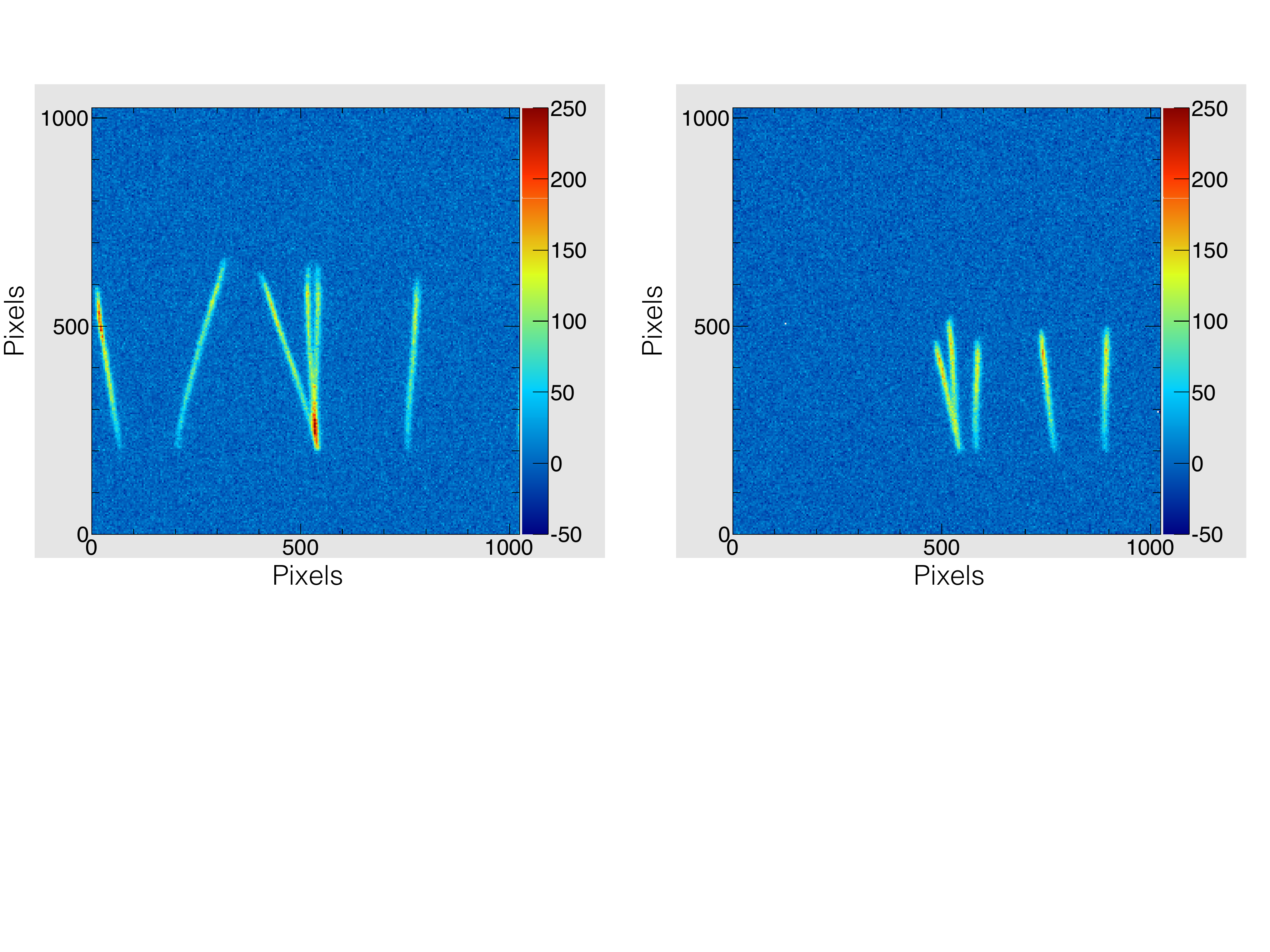} 
\vspace{-4cm}
\caption{$^{241}$Am alpha tracks in the $^4$He (left) and Ne mixtures (right). The Z-axis is indicative of the ``CCD weight'' or brightness. The Bragg peaks, indicative of the alphas' direction, are clearly visible. The resolution in the prototype detector is 0.16~mm/pixel.  \label{HeNetracks}}
\end{center}
\end{figure}

\subsection{Reconstruction of High Energy Neutrons}

Figure~\ref{HeNetracks} shows CCD readout images of characteristic tracks in both the $^4$He and Ne gas mixtures. It was immediately apparent that, in the Ne gas mixture, the track length was shorter and the slightly increased electron transverse diffusion would not compromise event reconstruction. The gain required to fully reconstruct events is also acceptable. Each image has multiple tracks because the alpha source was highly active and events often pile-up during readout\footnote{The $^{241}$Am source was very active, so while the camera exposure was 20~ms, its minimum value, multiple tracks occurred during each exposure. Due to a lack of shutter, the CCD remained exposed during readout. Events captured during readout can appear shifted horizontally, but actually originated at the source \cite{prototype}.}.

\begin{figure}[tb]
\begin{center}
\includegraphics[scale=0.375]{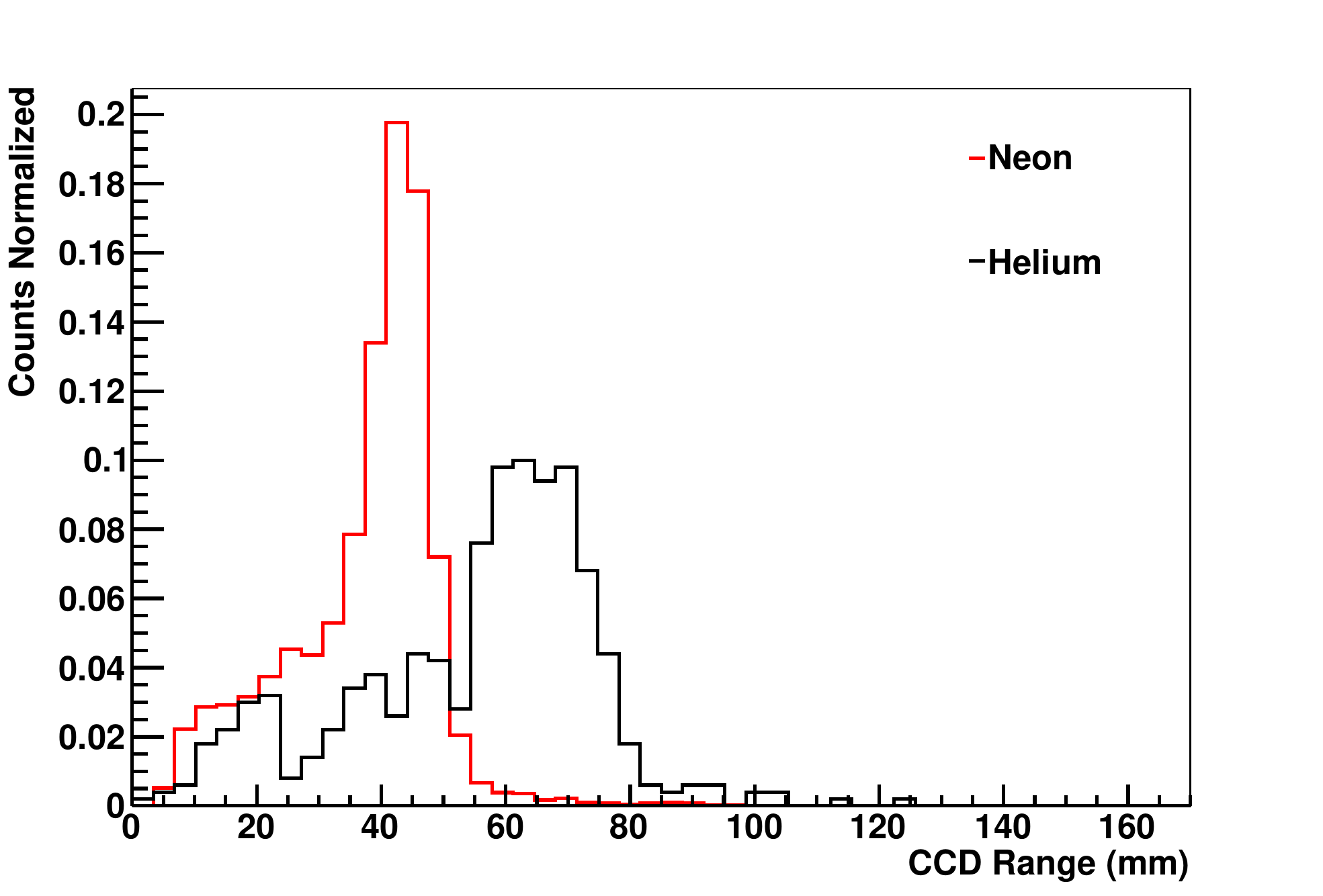}
\includegraphics[scale=0.375]{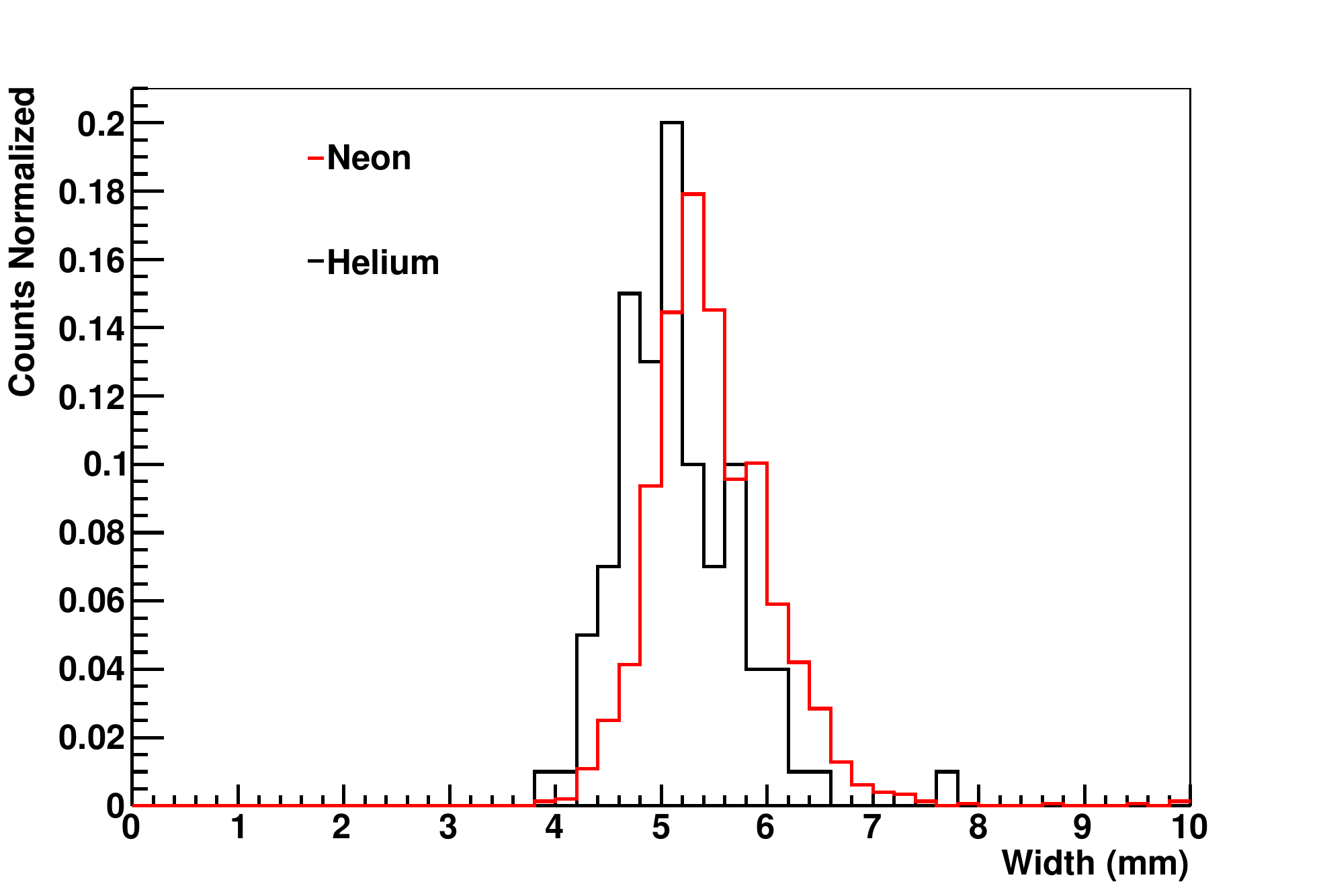}
\caption{Left: the CCD range of tracks in the $^4$He and Ne gas mixtures. Right: the characteristic track widths in the two mixtures. \label{Neccdfit}}
\end{center}
\end{figure}

As anticipated, Figure~\ref{Neccdfit} shows the significantly shorter CCD range of tracks in Ne compared to tracks in $^4$He. The average track length in the $^4$He mixture is 52$\%$ greater than in the Ne mixture. This, compounded with the fact that neutron-induced Ne recoils are significantly shorter than neutron-induced $^4$He recoils, will allow MITPC to fully contain and reconstruct more energetic nuclear recoils ($\lesssim$100 MeV). The track range as a function of kinetic energy for the relevant nuclei in the gas mixtures can be seen in Fig.~\ref{srim}. The characteristic energy loss as a function of distance can also be considered more desirable in the Ne mixture, as compared to the $^4$He mixture, given that radiogenic alpha backgrounds can contribute significantly to event rate. Radiogenic alphas and neutron-induced $^4$He recoils have identical energy loss profiles.

The transverse diffusion in pure Ne is expected to be about 4~times greater than that in $^4$He, given MITPC's typical drift field of $\sim$150~V/cm and pressure of 600~torr~\cite{noble_gas}, although the addition of CF$_4$ will reduce the size of the effect. As shown in Figure~\ref{Neccdfit}, we found that the track width, related to transverse diffusion, in the Ne mixture (given a drift distance of 6.2~cm) was about 9\% larger than with the $^4$He mixture\footnote{This plot was created with a cut on the CCD range of tracks, eliminating the low energy tails seen in the left plot in Figure~\ref{Neccdfit}. This ensured that the tracks considered for the width measurement were parallel to the amplification plane, and therefore consistent with each other in terms of range and drift distance.}. In consideration of the full drift distance in MITPC (32~cm), the increased width neither compromises event reconstruction nor reduces the signal-to-noise ratio below an acceptable level.

\subsection{Spark Rate}

Voltage breakdowns (sparks) in the amplification plane occur frequently because the anode plate and ground mesh are 0.440~mm apart, with a potential difference of $\sim$700~V between them. This high electric field is necessary in order to achieve the gain ($\sim$$10^5$) required for viewing tracks with the CCD camera. The minimum signal-to-noise requirement for maximizing the CCD track reconstruction abilities is $\approx$10:1. After a spark, data collection ceases for about 10~seconds in order to allow the amplification plane conductor voltages to return to normal conditions. Lower spark rates are therefore desirable because they reduce downtime. 

The spark rate with Ne mixture was found to be negligible. This can be compared to the $\sim 20\%$ downtime due to sparks experienced with the $^4$He mixture. The reason for the lower spark rate is twofold. First, we were able to obtain sufficient gain with Ne while running at a lower voltage than we could with $^4$He (680~V vs. 710~V). Second, Paschen's law (see, e.g., Ref.~\cite{paschen}) shows that the breakdown voltage for Ne is higher than that of $^4$He at these operating conditions.

\section{Conclusion}

In conclusion, we find that filling the MITPC detector with a gas composition of 87.5$\%$ Ne and 12.5$\%$ CF$_4$ at 600~torr will allow the energy and directional reconstruction of neutron events up to 100 MeV and possibly higher. This is the nuclear recoil energy range expected for neutrino- and beam-induced neutrons at the BNB at FNAL. We find that the diffusion of electrons and the signal-to-noise observed in the Ne mixture is acceptable, and will not compromise event reconstruction. The spark rate in the Ne mixture was found to be negligible, and will improve the detector uptime by about 20$\%$, as compared to $^4$He running. The success of these tests serves as a demonstration that Ne is an optimal fast-neutron target for characterizing the energy and direction of 10s-of-MeV neutrons using the TPC technology described. MITPC will begin running in the Ne configuration with beam data in October, 2015.

\section*{Acknowledgments}

We have collaborated closely with J. Dawson and A. Houlier on the construction, running, and analysis of DCTPC. This work, to be published, has informed this result. We also thank Z. Moss and B.J.P. Jones for carefully reading this paper.  The authors were supported by NSF-PHY-1505855.

\end{document}